\documentclass[letterpaper, 10 pt, conference]{ieeeconf}  

\IEEEoverridecommandlockouts                              
\usepackage{tikz}
\usepackage{amsmath}
\usepackage{latexsym, amssymb, amsmath, amsbsy,amsopn, amstext}
\usepackage{graphicx,hyperref,booktabs}
\usepackage{cancel, color}
\usepackage{epstopdf,subcaption}
\graphicspath{{images/}}

\DeclareMathOperator{\Col}{Col}
\DeclareMathOperator{\Row}{Row}
\DeclareMathOperator{\tr}{trace}

\def\cal{\mathcal}

\def\ra{\rightarrow}

\def\lra{\leftrightarrow}

\def\d{\delta}

\def\D{\Delta}

\def\0{{\bf 0}}

\def\con{con}

\newcommand{\N}{{\mathbb N}}

\def\dsum{\mathop{\sum}\limits}

\newtheorem{thm}{Theorem}[section]
\newtheorem{dfn}[thm]{Definition}
\newtheorem{prp}[thm]{Proposition}
\newtheorem{exa}[thm]{Example}
\newtheorem{rem}[thm]{Remark}
\newtheorem{cor}[thm]{Corollary}

\title{\LARGE \bf
	Transition System Representation of Boolean Control Networks
}

\author{Daizhan Cheng, Xiao Zhang and Zhengping Ji
	\thanks{This work is supported partly by NNSF 62073315 of China, and China Postdoctoral Science Foundation 2021M703423 and 2022T150686.}
\thanks{D. Cheng is with the Academy of Mathematics and Systems Science, Chinese Academy of Sciences, Beijing 100190, P.R.China, {\tt\small dcheng@iss.ac.cn}}
\thanks{X. Zhang is with the National Center for Mathematics and Interdisciplinary Sciences \& the Key Laboratory of Systems and Control, Academy of Mathematics and Systems Science, Chinese Academy of Sciences, Beijing 100190, P.R.China, {\tt\small xiaozhang@amss.ac.cn}}
\thanks{Z. Ji is with the Key Laboratory of Systems and Control, Academy of Mathematics and Systems Science \& School of Mathematical Sciences, University of Chinese Academy of Sciences, Beijing 100190, P.R.China, {\tt\small jizhengping@amss.ac.cn} }
}

\begin{document}

	\maketitle
	\thispagestyle{empty}
	\pagestyle{empty}

\begin{abstract}
	
First, the topological structure of a transition system is studied. Then, two types of transition system (TS) representations of Boolean networks (BNs) and Boolean control networks (BCNs) are investigated. The first kind of representation is state-based, which converts a BCN into a TS with either distinct control or non-distinct control. The second representation is output-based, which is also called the simulation of the original BCN. Some applications are also studied.	
\end{abstract}

\section{Introduction}

Since Kauffman proposed the BN to formulate genetic networks \cite{kau69}, the study of BNs and BCNs becomes a heat topic in the biological community as well as in the control community.
Consider a BN,
\begin{align}\label{1.1}
\begin{cases}
X_1(t+1)=f_1(X_1(t),X_2(t),\cdots,X_n(t)),\\
X_2(t+1)=f_2(X_1(t),X_2(t),\cdots,X_n(t)),\\
\vdots,\\
X_n(t+1)=f_n(X_1(t),X_2(t),\cdots,X_n(t)),\\
\end{cases}
\end{align}
where $X_i\in {\cal D}:=\{0,1\}$,  $f_i:{\cal D}^n\ra {\cal D}$, $i\in [1,n]$.

It is clear that every trajectory can converge to an attractor (either a fixed point or a cycle) because it consists only of finite nodes and each node can take only two values ${\cal D}=\{0,1\}$. Thus, the attractors, with their attractor basins, form the entire topological structure of a BN. Therefore, finding both fixed points and cycles of a given BN becomes a first priority problem in the study of BN. Many early works considered this problem by providing methods to solve a certain class of BNs \cite{far04,hua00,goo63,hei03}, to name a few.

In the last two decades, the semi-tensor product (STP) of matrices has been used to transform BN (or BCN) into an algebraic discrete-time dynamical (control) system. The STP approach to BN (BCN) has proven to be very efficient. Many theoretical results have been obtained.

The basic step for the STP approach can be described in a nutshell as follows: construct the so-called vector form of $X_i$ as
$$
x_i:=\begin{bmatrix}
X_i,\\
1-X_i,
\end{bmatrix} \in \D,\quad i\in [1,n],
$$
where $\D:=\D_2$, and $\D_k:=\Col(I_k)$ is the column set of the identity matrix $I_k$.

Using the vector form for $X_i$, the system (\ref{1.1}) can be expressed in its algebraic state space representation (ASSR) as \cite{che11}
\begin{align}\label{1.2}
x(t+1)=Mx(t),
\end{align}
where $x(t)=\ltimes_{i=1}^nx_i(t)\in \D_{2^n}$ and $M=M_1*M_2*\cdots,*M_n \in {\cal L}_{2^n\times 2^n}$, with $M_i$ being the structure matrix of $f_i$, $*$ is the Khatri-Rao product of matrices \cite{che12}.

The general formula for calculating the number of attractors is given by \cite{che10} as

\begin{thm}\label{t1.1} \cite{che10} Consider the Boolean network (\ref{1.1}) with its ASSR (\ref{1.2}). Then
\begin{align}\label{1.3}
\begin{cases}
N_1=\tr(M),\\
N_s=\frac{\tr(M^s)-\dsum_{k\in {\cal P}(s)}kN_k}{s},\quad 2\leq s\leq n,
\end{cases}
\end{align}
where $N_s$ is the number of cycles of length $s$, in particular, $N_1$ is the number of fixed points considered as cycles of length $1$. Note that ${\cal P}(s)$ is the set of the proper factors of $s$, including $1$ and excluding $s$.
\end{thm}

\begin{rem}\label{r1.2} The proof of Theorem \ref{t1.1} is based on the following three observations:
\begin{itemize}
\item[(i)] If $x_i\in\D_{2^n}$ is on a cycle of length $s$, then $x_i$ is a fixed point of $M^s$.
\item[(ii)] If $x_i\in\D_{2^n}$ is on a cycle of length $t$ and {$t|s$}, then $x_i$ is also a fixed point of $M^s$.
\item[(iii)] If $x_i\in\D_{2^n}$ is a fixed point of $M^s$, it must be either of type (i) or of type (ii).
\end{itemize}
\end{rem}

 Consider a $k$-valued logical network. Assume it is expressed as in (\ref{1.1}) with only $X_i\in {\cal D}_k=\{1,\cdots,k\}$, $i\in [1,n]$. Let $j\sim \d_k^j$, $j\in [1,k]$, and
$$
x_i=\d_k^j,\quad X_i=j,\quad i\in [1,n],\; j\in [1,k],
$$
is the vector form of $X_i$. Then its ASSR is still expressed as in (\ref{1.2}) with only $M\in {\cal L}_{k^n\times k^n}$.

Taking into account the observations of Remark \ref{r1.2}, the following result is also obvious.

\begin{cor}\label{c1.3} \cite{li10} Consider the $k$-valued logical network (\ref{1.1}) with its ASSR (\ref{1.2}). Then
the formula (\ref{1.3}) remains true.
\end{cor}

The cycles of BCN have also been studied in several papers, e.g. \cite{zha10,las13} etc. But there is no formula similar to (\ref{1.3}) to calculate all control attractors.

A transition system (TS) is a more general finite-valued network. A BN can be seen as an autonomous TS and a BCN as a TS.
The TS itself is a very useful framework for representing finite automata (FA) \cite{gua87,kar05,kar13}. In particular, it provides a fundamental framework for hybrid systems \cite{tab09,lin14}. The STP approach to FAs has also been developed \cite{xu13,xu13b,den22}.

To our best knowledge, the topological structure of a TS, or the structure of its attractors, has not been clearly revealed. ``Can the formula (\ref{1.3}) be applied to them?" is still an open problem.

In this paper, we first consider the topological structure of autonomous TS. Two types of cycles are considered: simple cycles and compound cycles. Then the number of attractors with different lengths is calculated. Then the transformation of BCN into autonomous TS is considered. Using transformed TS, the attractors of BCN can be calculated. Finally, some applications of such transformations are examined.

The remainder of this paper is structured as follows:

Section II considers the topological structure of TSs, and the formula of fixed points and cycles for networks is extended to TSs. Section III investigates the state-based representation of BCNs with some direct applications. Section IV discusses the output-based representation of BCNs, called the simulation of BCNs. Furthermore, the formula for the simulation dynamics is obtained. The output robust controls are also investigated. Section V is a brief conclusion.

The notations used in the text are listed below:
\begin{itemize}
    \item $\Col(A)$ ($\Row(A)$): the set of columns (rows) of $A$.
    \item ${\cal D}_k:=\{1,2,3,\cdots,k\}$.
    \item $\delta_n^i$: the $i$-th column of the identity matrix $I_n$.
    \item $\Delta_n:=\Delta_n=\{\d_n^i\,|\,i=1,2,\cdots,n\}$.
    \item ${\cal L}_{m\times n}$: the set of ${m\times n}$ logical matrices, that is, $\Col({\cal L})\subset \Delta_{m}$.
    \item ${\cal B}_{m\times n}$: the set of $m\times n$ Boolean matrices, that is, $[{\cal B}]_{i,j}\in {\cal D}:=\{0,1\}$.
    \item $\d_{m}[i_1,i_2,\cdots,i_n]:=[\d_m^{i_1},\d_m^{i_2},\cdots,\d_m^{i_n}]\in {\cal L}_{m\times n}$.
  \item $A+_{\cal B}B$: the Boolean addition of $A,B\in{\cal B}_{m\times n}$ that is, $[A+_{\cal B}B]_{i,j}=[A]_{i,j}\vee [B]_{i,j}$, $\dsum_{\cal B}$ is the Boolean sum.
   \item $A\times_{\cal B}B$: the Boolean product of $A,B$. For $A\in {\cal B}_{n\times n}$, $A^{(s)}:=\underbrace{A\times_{\cal B}A\times_{\cal B}\cdots\times_{\cal B}A}_{s}$.
\end{itemize}

\section{Transition Systems}

\begin{dfn}\label{d2.1} \cite{bel17} A TS can be described by  $T=({\cal X},{\cal U},\Sigma,{\cal O},h)$, where
\begin{itemize}
\item[(i)] ${\cal X}:=\{X_1,X_2,\cdots,X_n\}$ is a finite state set and $X_i\in {\cal D}_2$ for Boolean TS (or $X_i\in {\cal D}_k$ for $k$-valued TS);
\item[(ii)] ${\cal U}:=\{U_1,U_2,\cdots,U_m\}$ is a finite input set and $U_j\in {\cal D}_2$ for Boolean TS (or $U_j\in {\cal D}_k$ for $k$-valued TS);
\item[(iii)] $\Sigma:{\cal X}\times {\cal U} \ra 2^{\cal X}$ is the state transition mapping;
\item[(iv)] ${\cal O}:=\{O_1,O_2,\cdots, O_p\}$ is the observing set;
\item[(v)] $h:{\cal X}\ra {\cal O}$ is the observation mapping.
\end{itemize}
If $|\Sigma(X, U)|\leq 1$, for $\forall X\in {\cal X}$ and $\forall U\in {\cal U}$, then $T$ is called a deterministic TS, otherwise, it is called a non-deterministic TS.
\end{dfn}

Then the dynamics of a TS can be expressed as
\begin{align}\label{2.1}
\begin{cases}
X(t+1)=\Sigma(U(t),X(t)),\\
Y(t)=h(X(t)).
\end{cases}
\end{align}

For a TS where $|{\cal X}|=n$, $|{\cal U}|=m$, and $|{\cal O}|=p$,
a subset $X(t)\in 2^{\cal X}$ can be expressed into vector form as $x(t)=(x_1(t),x_2(t),\cdots,x_n(t))^\mathrm{T} \in {\cal B}^n$, which is a Boolean vector, where
$$
x_i(t)=
\begin{cases}
1,\quad X_i\in X(t),\\
0,\quad X_i\not\in X(t),\quad i\in[1,n].
\end{cases}
$$

Using vector form expressions, similar to BN (or $k$-valued network), the ASSR of a transition system can be expressed as
\begin{align}\label{2.2}
\begin{cases}
x(t+1)=Lu(t)x(t),\\
y(t)=Hx(t),\\
\end{cases}
\end{align}
where $L\in {\cal B}_{n\times nm}$ is a Boolean matrix, $H\in {\cal L}_{p\times n}$ is a logical matrix.

It is also clear that the ASSR of an autonomous TS is
\begin{align}\label{2.3}
\begin{cases}
x(t+1)=Mx(t),\\
y(t)=Hx(t),\\
\end{cases}
\end{align}
where, $M\in {\cal B}_{n\times n}$ is a Boolean matrix.

\begin{exa}\label{efm.1.2} \cite{bel17} Consider a TS as
$T=({\cal X},{\cal U},\Sigma,{\cal O},h)$ (Ref. to Fig. \ref{Fig.fm.1.1}), where
\begin{itemize}
\item[(i)] ${\cal X}=\{x_1,x_2,x_3,x_4\}$.
\item[(ii)] ${\cal U}=\{u_1,u_2\}$.
\item[(iii)]
$$
\begin{array}{ll}
\Sigma(x_1,u_1)=\{x_2,x_3\},&\Sigma(x_2,u_1)=\{x_2,x_3\},\\
\Sigma(x_2,u_2)=\{x_4\},&\Sigma(x_3,u_2)=\{x_2,x_3\},\\
\Sigma(x_4,u_1)=\{x_2,x_4\}.&~\\
\end{array}
$$
\item[(iv)] ${\cal O}=\{O_1,O_2,O_3\}$.
\item[(v)]
$
h(x_1)=O_1,\quad h(x_2)=h(x_4)=O_2,\quad h(x_3)=O_3.
$
\end{itemize}

\vskip 5mm

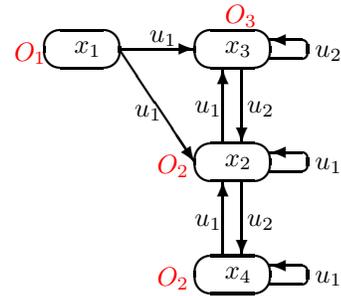
\begin{figure}
\centering
\setlength{\unitlength}{0.5cm}
\begin{picture}(9,9)\thicklines
\put(2,7.5){\oval(2,1)}
\put(6,1.5){\oval(2,1)}
\put(6,4.5){\oval(2,1)}
\put(6,7.5){\oval(2,1)}
\put(3,7.5){\vector(1,0){2}}
\put(3,7.5){\vector(2,-3){2}}
\put(5.75,5){\vector(0,1){2}}
\put(6.25,7){\vector(0,-1){2}}
\put(5.75,2){\vector(0,1){2}}
\put(6.25,4){\vector(0,-1){2}}
\put(7,1.5){\oval(2,0.5)[r]}
\put(7.5,1.75){\vector(-1,0){0.5}}
\put(7,4.5){\oval(2,0.5)[r]}
\put(7.5,4.75){\vector(-1,0){0.5}}
\put(7,7.5){\oval(2,0.5)[r]}
\put(7.5,7.75){\vector(-1,0){0.5}}
\put(1.8,7.4){$x_1$}
\put(5.8,7.4){$x_3$}
\put(5.8,4.4){$x_2$}
\put(5.8,1.4){$x_4$}
\put(3.8,7.7){$u_1$}
\put(3.4,5.7){$u_1$}
\put(5,5.8){$u_1$}
\put(5,2.8){$u_1$}
\put(6.4,5.8){$u_2$}
\put(6.4,2.8){$u_2$}
\put(8.2,7.3){$u_2$}
\put(8.2,4.3){$u_1$}
\put(8.2,1.3){$u_1$}
\put(0.2,7.2){${\color{red}O_1}$}
\put(5.8,8.2){${\color{red}O_3}$}
\put(4,4.2){${\color{red}O_2}$}
\put(4,1.2){${\color{red}O_2}$}
\end{picture}
\caption{TS of Example \ref{efm.1.2} \label{Fig.fm.1.1}}
\end{figure}

Let
$$
\begin{array}{ll}
x_i=\d_4^i,& i=1,2,3,4;\\
u_j=\d_2^j,&j=1,2;\\
y_k=\d_3^k,& k=1,2,3.
\end{array}
$$
Then the ASSR of $T$ is
\begin{align}\label{fm.1.2}
\begin{cases}
x(t+1)=Lu(t)x(t),\\
y(t)=Hx(t),
\end{cases}
\end{align}
where,
$$
L=\begin{bmatrix}
0&0&0&0&0&0&0&0\\
1&1&0&1&0&0&1&0\\
1&1&0&0&0&0&1&0\\
0&0&0&1&0&1&0&0\\
\end{bmatrix},
$$
$$
H=\d_3[1,2,3,2].
$$
\end{exa}

\begin{dfn}[\cite{bel17}]\label{d2.2}  Consider the transition system (\ref{2.1}).
\begin{itemize}
\item[(i)] A set $X(t,U,X(0)):=\{X(0),X(1),\cdots,X(t)\}$ is called a trajectory starting from $X(0)$ and driven by $U=\{U(0),U(1),\cdots,U(t-1)\}$, where $X(\tau)\in {\cal X}$, $\tau\in[0,t]$, and
$$
X(\tau+1)\in \Sigma(U(\tau),X(\tau)),\quad \tau\in[0,t-1].
$$
\item[(ii)] A trajectory $\{X(\tau),X(\tau+1),\cdots,X(\tau+\ell-1)\}$ is called a general cycle (GC) of length $\ell$ if $X(\tau+\ell)=X(\tau)$.

\item[(iii)] A cycle of length $1$ is called a fixed point.

\item[(iv)] A cycle is called a simple cycle (SC) if $X(\tau)$, $X(\tau+1)$, $\cdots$, $X(\tau+\ell-1)$ are distinct (fixed points are also considered SCs). Otherwise, it is called a compound cycle (CC).

\item[(v)] A cycle $C_P$ is called a (non-trivial) power cycle (PC), if $C$ is a cycle and $C_P=\underbrace{C~C~\cdots~C}_k$, $k\geq 2$. Obviously, a PC must be a CC.
\end{itemize}
\end{dfn}

\begin{rem}\label{r2.3} The cycle of an autonomous TS can be defined as a simplified version of Definition \ref{d2.2}. We omit it and assume that the cycle of an autonomous TS is also properly defined.
\end{rem}

\begin{exa}\label{e2.4} Consider an autonomous TS, denoted by $T$, which has a transition graph as shown in Fig. \ref{Fig.2.1}.
\begin{figure}
\centering
\setlength{\unitlength}{1 cm}
\begin{picture}(6,3)(-0.5,-0.5)\thicklines
\put(1.5,1){\oval(1,2)}
\put(4.5,1){\oval(1,2)}
\put(2,1.5){\vector(1,0){2}}
\put(4,0.5){\vector(-1,0){2}}
\put(1,1){\oval(1,1)[l]}
\put(0.75,0.5){\vector(1,0){0.25}}
\put(1.25,1){$x_1$}
\put(4.25,1){$x_2$}
\end{picture}
\caption{Transition System $T$ \label{Fig.2.1}}
\end{figure}
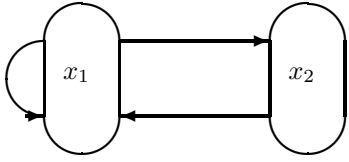
Its ASSR is easily obtained as
\begin{align}\label{2.4}
x(t+1)=\begin{bmatrix}
1&1\\
1&0
\end{bmatrix}
x(t).
\end{align}
Then
\begin{itemize}
\item[(i)] $1$ is a fixed point.
\item[(ii)] $(1,2)$ is an SC.
\item[(iii)] $(1,2,1,2,1,2)$ is a PC.
\item[(iv)]
$
(1,2,1,1,2,1,1,1,2,\cdots, \underbrace{1,\cdots,1}_s, 2),
\quad s=1,2,\cdots
$
are CCs of length $\frac{s(s+3)}{2}$, which can be arbitrarily long.
\end{itemize}
\end{exa}

Consider a partition as:
$$
\mbox{GC}~=~\mbox{SC}~\bigcup ~\mbox{CC}.
$$

Following the argument in Remark \ref{r1.2}, it is easy to obtain the following result:

\begin{prp}\label{p2.5} Consider an autonomous TS with its ASSR (\ref{2.3}). Then formula (\ref{1.3}) is applicable to calculate its numbers of CCs of arbitrary length $s>0$.
\end{prp}

As shown in Example \ref{e2.4}, the length $s>0$ of a CC could be arbitrarily large, so applying formula (\ref{1.3}) to calculate the numbers of all its CCs with arbitrary length $s>0$ is impossible. The problem can be solved by observing the following facts:

\begin{itemize}
    \item An SC is a cycle of length $s\leqslant n$;
    \item Any CC can be viewed as a recursive concatenation or insertion of SCs. For example, consider a trajectory of the form $C:=(y,\cdots,\xi,x_1,x_2,\cdots,x_{\ell-1}, \xi,\cdots,y)$ where $x_1,x_2,\cdots,x_{\ell-1}$ are distinct; then $(\xi,x_1,x_2,\cdots,x_{\ell-1})$ is an SC (of length less than $n$); replace the subsequence $\xi,x_1,x_2,\cdots,x_{\ell-1}, \xi$ in $C$ by $\xi$ and keep finding subcycles that are SCs in the remaining trajectory, we will end up partitioning the trajectory into SCs and finding it constructed by recursively inserting SCs into an SC.
\end{itemize}

Therefore, we only have to compute all SCs using the formula (\ref{1.3}) and the algorithm for finding attractors of BNs (or $k$-valued networks) \cite{che10}. Then the SCs are sufficient to describe the topological structure of the autonomous TS.

We consider a simple example.

\begin{exa}\label{e2.6} Consider a TS, its ASSR is
\begin{align}\label{2.5}
x(t+1)=\begin{bmatrix}
0&0&1&0\\
1&0&0&0\\
1&1&0&0\\
0&0&1&1\\
\end{bmatrix}
x(t):=Mx(t).
\end{align}
Using the formula (\ref{1.3}), it is easy to calculate that
$$
N_1=1,\quad N_2=1,\quad N_3=1,\quad N_4=0.
$$
The corresponding attractors are:
$$
(4),\quad
(1,3),\quad
(1,2,3).
$$
They are all SCs.
\end{exa}

\section{TS Representation of BCNs}

\subsection{Conversion of BCNs to Autonomous TSs}

Consider a BCN as
\begin{align}\label{3.1}
\begin{cases}
X_1(t+1)=f_1(X_1(t),\cdots,X_n(t),U_1(t),\cdots,U_m(t)),\\
X_2(t+1)=f_2(X_1(t),\cdots,X_n(t),U_1(t),\cdots,U_m(t)),\\
\vdots\\
X_n(t+1)=f_n(X_1(t),\cdots,X_n(t),U_1(t),\cdots,U_m(t)),\\
\end{cases}
\end{align}
where $X_i,U_j\in {\cal D}$,  $f_i:{\cal D}^{n+m}\ra {\cal D}$, $i\in [1,n]$, $j\in [1,m]$.

The ASSR of (\ref{3.1}) is
\begin{align}\label{3.2}
x(t+1)=Lu(t)x(t),
\end{align}
where $x(t)=\ltimes_{i=1}^nx_i(t)\in\Delta_{2^n}$, $u(t)=\ltimes_{j=1}^mu_j(t)\in\Delta_{2^m}$, $L\in {\cal L}_{2^n\times 2^{m+n}}$.

\begin{dfn}\label{d3.1}
Consider system (\ref{3.1}).
\begin{itemize}
\item[(i)]
If there exists a sequence $$u(\tau),u(\tau+1),\cdots,u(\tau+s-1)$$ such that
$$
\begin{array}{l}
x(\tau)\xrightarrow{u(\tau)} x(\tau+1)\xrightarrow{u(\tau+1)} x(\tau+2)\cdots x(\tau+s-1)\\
\xrightarrow{u(\tau+s-1)}
x(\tau+s).
\end{array}
$$
Then $
x(\tau),x(\tau+1),\cdots, x(\tau+s)$ is called a control trajectory with undistinguished control of length $s$.

The sequence of state-control pairs $
(x(\tau), u(\tau)) ,(x(\tau+1),u(\tau+1)),\cdots, (x(\tau+s),u(\tau+s)) $ is called a control trajectory with distinguished control of length $s$.

\item[(ii)]
A control trajectory $x(\tau),x(\tau+1),\cdots, x(\tau+s)$ is called a control cycle of length $s$, if $x(\tau)=x(\tau+s)$.
The simple (or power, compound, etc.) control cycle can be defined similarly.

\item[(iii)] A control cycle of length $1$ is called a control fixed point.
\end{itemize}
\end{dfn}

\begin{dfn}\label{d3.2} Consider BCN (\ref{3.1}).
\begin{itemize}
\item[(i)] It is converted into an autonomous TS with  undistinguished control, where the ASSR of the TS is
\begin{align}\label{3.3}
x(t+1)=Mx(t),
\end{align}
where
\begin{align}\label{3.4}
M={\dsum_{{\cal B}}}_{i=1}^{2^m}(L\d_{2^m}^i),
\end{align}
here $\dsum_{{\cal B}}$ is the Boolean sum.

\item[(ii)] It is converted into an autonomous TS with  distinguished control, where the ASSR of the TS is
\begin{align}\label{3.5}
w(t+1)=\Xi w(t),
\end{align}
where $w(t)=u(t)x(t)$,
$
\Xi={\underbrace{[L^{\mathrm{T}}~L^{\mathrm{T}}\cdots~L^{\mathrm{T}}]}_{2^m}}^{\mathrm{T}}.
$
\end{itemize}
\end{dfn}

Using Proposition \ref{p2.5} to the converted autonomous TS yields the following result, which can be used to calculate control cycles of BCNs.

\begin{cor}\label{c3.3}
\begin{itemize}
\item[(i)] Applying the formula (\ref{1.3}) to the converted autonomous TS with undistinguished control (\ref{3.3}), the number of cycles
of a BCN with undistinguished control of different lengths $s$ can be calculated.

\item[(ii)] Applying the formula (\ref{1.3}) to the converted autonomous TS with  distinguished control (\ref{3.4}), the number of cycles of a BCN with distinguished control of different lengths $s$ can be calculated.
\end{itemize}
\end{cor}

\begin{rem}\label{r3.4} With some mild revision, the above results can be naturally extended to $k$-valued control networks. Similarly, they are also applicable to a general TS, when it is converted into an autonomous TS. The following example shows this.
\end{rem}

\begin{exa}\label{e3.5}  Recall Example \ref{efm.1.2}.
\begin{itemize}

\item[(i)] A straightforward computation shows that its converted autonomous TS with undistinguished control is determined by its ASSR as
$$
x(t+1)=M_Ix(t),
$$
where the transition matrix is
\begin{align}\label{fm.6.9}
M_I=\begin{bmatrix}
0&0&0&0\\
1&1&1&1\\
1&1&1&0\\
0&1&0&1\\
\end{bmatrix}.
\end{align}

\item[(ii)]  The ASSR of its converted autonomous TS with distinguished control is
$$
z(t+1)=\Xi z(t),
$$
where
\begin{align}\label{fm.6.10}
\Xi=\begin{bmatrix}
0&0&0&0&0&0&0&0\\
1&1&0&1&0&0&1&0\\
1&1&0&0&0&0&1&0\\
0&0&0&1&0&1&0&0\\
0&0&0&0&0&0&0&0\\
1&1&0&1&0&0&1&0\\
1&1&0&0&0&0&1&0\\
0&0&0&1&0&1&0&0\\
\end{bmatrix}.
\end{align}

\item[(iii)]
Using the formula (\ref{1.3}) the number of CCs with undistinguished control are
\begin{itemize}
\item[(a)] $N_1=3$:
$$
X_{\mbox{fixed point}}=\{2,3,4\},
$$
where, $i$ stands for $x_i$, $i=1,2,3,4$.
\item[(b)] $N_2=2$:
$$
C^2_1=(2,3);\quad C^2_2=(2,4).
$$
\item[(c)] $N_3=4$:
$$
\begin{array}{ll}
C^3_1=(2,2,3);& C^3_2=(2,3,3);\\
C^3_3=(2,2,4);& C^3_4=(2,4,4).
\end{array}
$$
\item[(d)] $N_4=7$:
$$
\begin{array}{ll}
C^4_1=(2,2,2,3);& C^4_2=(2,2,3,3);\\
C^4_3=(2,3,3,3);& C^4_4=(2,2,2,4);\\
C^4_5=(2,2,4,4);& C^4_6=(2,4,4,4);\\
C^4_7=(2,3,2,4).& ~\\
\end{array}
$$
\item[(e)] $N_5=16$:
$$
\begin{array}{ll}
C^5_1=(2,2,2,2,3);& C^5_2=(2,2,2,3,3);\\
C^5_3=(2,2,3,3,3);& C^5_4=(2,3,3,3,3);\\
C^5_5=(2,3,2,2,3);& C^5_6=(2,3,2,3,3);\\
C^5_7=(2,2,2,2,4);& C^5_8=(2,2,2,4,4);\\
C^5_9=(2,2,4,4,4);& C^5_{10}=(2,4,4,4,4);\\
C^5_{11}=(2,4,2,2,4);& C^5_{12}=(2,4,2,4,4);\\
C^5_{13}=(2,3,2,2,4);& C^5_{14}=(2,3,2,4,2);\\
C^5_{15}=(2,3,3,2,4);& C^5_{16}=(2,4,4,2,3).\\
\end{array}
$$
$$
\vdots
$$

Finally, the SC is
$$
SC:\{\{2\},\{3\}, \{4\}, \{2,3\}, \{2,4\}\}.
$$

\end{itemize}
\end{itemize}
\end{exa}

\subsection{Some Applications}

\begin{itemize}
\item Reachability of TSs:

Consider an autonomous TS (in ASSR form):
\begin{align}\label{4.1.1}
x(t+1)=Mx(t),
\end{align}
where $x(t)\in \D_n$, $M\in {\cal B}_{n\times n}$.

\begin{dfn}\label{d4.1.1} $x_j$ is reachable from $x_i$, if the trajectory $x(t,x_0)$, starting from $x_0=x_i$, can reach $x_j$ at finite time $t_0$, i.e., $x(t_0,x_0)=x_j$.
\end{dfn}

Define the reachable matrix ${\cal C}$ as
\begin{align}\label{4.1.2}
{\cal C}={\dsum_{{\cal B}}}_{s=1}^n M^{(s)}.
\end{align}

Then the following result is well known.

\begin{prp}[\cite{che11}] \label{p4.1.2} Assume (\ref{4.1.1}) is the converted autonomous TS of a BCN $\Sigma$. Then for (\ref{4.1.1}) $x_j$ is reachable from $x_i$, if and only if, for BCN $\Sigma$: $x_j$ is reachable from $x_i$.
\end{prp}

\item Decoupling:

\begin{dfn}\label{d4.1.3}
\begin{itemize}
\item[(i)] A subset $Z\subset \D_n$ is called an attractor of (\ref{4.1.1}), if $x(t)\in Z$ implies $x(t+1)\in Z$ for $\forall t\in \N$.
\item[(ii)] Assume $Z=\{x_0\}$ is an attractor, then $Z$ (or $x_0$) is called a fixed point.
\end{itemize}
\end{dfn}

The following result is obvious:
\begin{prp}\label{p4.1.3}
\begin{itemize}
\item[(i)] Suppose (after a coordinate change if necessary) $Z=\{x_1,x_2,\cdots,x_r\}$. Then $Z$ is an attractor, if and only if
$M$ has the following block upper triangle form:
\begin{align}\label{4.1.3}
M=\begin{bmatrix}
M_1&M_2\\
0&M_3\\
\end{bmatrix}
\end{align}
where $M_1\in {\cal B}_{r\times r}$.

\item[(ii)] Suppose (after a possible coordinate change) $Z_i=\{x^i_1,x^i_2,\cdots,x^i_{r_i}\}$, $i=1,2,\cdots,s$ are sets of disjoint states. Then $Z_i$, $i=1,2,\cdots,s$ are attractor sets if and only if,
$M$ has the form (\ref{4.1.2}), where $M_1$ is a block diagonal matrix such that
$$
M_1=\begin{bmatrix}
M_1^1&0&\cdots&0\\
0&M_1^2&\cdots&0\\
\vdots&~&~&~\\
0&0&\cdots&M_1^s\\
\end{bmatrix},
$$
where $M_1^i\in {\cal B}_{r_i\times r_i}$, $i\in [1,s]$.
\end{itemize}
\end{prp}

\end{itemize}

\section{Simulation of BNs and BCNs}

\subsection{Output-Based Simulation}

Consider a logical control network or a deterministic TS (in ASSR form), denoted by $\Sigma$, and defined by
\begin{align}\label{5.1.1}
\begin{cases}
x(t+1)=Mu(t)x(t),\\
y(t)=Hx(t),
\end{cases}
\end{align}
where $x(t)\in \D_n$, $u(t)\in \D_m$, $M\in {\cal L}_{n\times mn}$, $H\in {\cal L}_{p\times n}$.

The following definition is based on \cite{bel17} with mild formation modification.

\begin{dfn}\label{d5.1.1} Consider  control network $\Sigma$ (\ref{5.1.1}).
\begin{itemize}
\item[(i)] Two states $x_i$ and $x_j$ are said to be (output) equivalent, denoted by $x_i\sim x_j$, if $Hx_i=Hx_j$.
\item[(ii)] The quotient system, denoted by $\Sigma/\sim$ is called a simulation of $\Sigma$.
\end{itemize}
\end{dfn}

Denote by $\bar{x}$ the equivalence class of $x$; $\con(\bar{x}):=\{x\;|\; x\sim \bar{x}\}$; ${\cal O}(x)$ the output trajectory for the state trajectory starting from $x$. Then we have

\begin{prp}\label{p5.1.2} \cite{bel17}
\begin{align}\label{5.1.2}
{\cal O}_{\Sigma}(\con(\bar{x}))\subset {\cal O}_{\Sigma/\sim}(\bar{x}).
\end{align}
\end{prp}

From the set controllability point of view \cite{guo15,che18}, the following is a straightforward result \cite{ji}.

\begin{prp}\label{p5.1.3} The simulation $\Sigma/\sim$ of $\Sigma$ is a transition system, and its dynamics is
\begin{align}\label{5.1.3}
\begin{cases}
\bar{x}(t+1)=\left[H\times_{{\cal B}} M \times_{{\cal B}} (I_m\otimes H^T)\right]u(t)\bar{x}(t),\\
y(t)=HH^T\bar{x}(t).
\end{cases}
\end{align}
\end{prp}

\subsection{Output-Robust Network and Control}

Consider a network (i.e., a deterministic TS):
\begin{align}\label{5.2.1}
\begin{cases}
x(t+1)=Mx(t),\\
y(t)=Hx(t),
\end{cases}
\end{align}
where $M\in {\cal L}_{n\times n}$, $H\in {\cal L}_{p\times n}$. Assume (\ref{5.2.1}) is a system with possible disturbances, described by
\begin{align}\label{5.2.2}
M=
\begin{cases}
M_0,\quad \xi\in \emptyset,\\
L\xi,\quad L\in {\cal L}_{n\times ns},~\xi\in \D_s.
\end{cases}
\end{align}
Then we have two models: when $M=M_0$ it is called the nominated model; and when $M=L\xi$ it is called the disturbed model.
For the nominated model, denoted by $\Sigma_0$, we can construct its simulation system $\Sigma_0/\sim$.
For the disturbed model, we can first get its TSR, denoted by $\Sigma_{\xi}$, and then construct its simulation system $\Sigma_{\xi}/\sim$.

\begin{dfn}\label{d5.2.1} System (\ref{5.2.1}) is said to be output robust, if $\Sigma_0/\sim$ and $\Sigma_{\xi}/\sim$ are {bi-simulated, that is to say, they generate identical output dynamics calculated from (\ref{5.1.3}).}
\end{dfn}

\begin{exa}\label{e5.2.2}
Consider a Boolean network $\Sigma$, which has its nominated model as
\begin{align}\label{5.2.3}
\begin{array}{l}
\begin{cases}
x_1(t+1)=\neg x_1(t),\\
x_2(t+1)=x_1(t)\bar{\vee} x_3(t),\\
x_3(t+1)=\left[x_1(t)\bar{\vee} x_2(t)\right]\vee x_3(t),
\end{cases}\\
~~y(t)=\left[x_1(t)\lra x_2(t)\right]\lra \neg x_3(t);
\end{array}
\end{align}
and its disturbed model as
\begin{align}\label{5.2.4}
\begin{array}{l}
\begin{cases}
x_1(t+1)=(\neg \xi(t))\wedge x_1(t),\\
x_2(t+1)=\left[\xi(t)\vee \neg x_1(t)\right]\bar{\vee} x_3(t),\\
x_3(t+1)=\left[x_1(t)\bar{\vee} x_2(t)\right]\vee x_3(t),
\end{cases}\\
~~y(t)=\left[x_1(t)\lra x_2(t)\right]\lra \neg x_3(t).
\end{array}
\end{align}

It is easy to have its ASSR as in (\ref{5.2.1}), where
$$
\begin{array}{l}
M_0=\d_{8}[7,6,7,5,1,3,1,4],\\
L=\d_{8}[7,6,7,5,7,6,1,4,1,3,7,5,7,6],\\
H=\d_2[2,1,1,2,1,2,2,1].
\end{array}
$$
The transition representation of its disturbed model becomes $x(t+1)=Tx(t)$, where
$$
T=\begin{bmatrix}
1&0&1&0&0&0&0&0\\
0&0&0&0&0&0&0&0\\
0&0&0&1&0&0&0&0\\
0&1&0&0&0&0&0&0\\
0&0&0&1&0&1&0&0\\
0&1&0&0&0&0&0&1\\
1&0&1&0&1&0&1&0\\
0&0&0&0&0&0&0&0\\
\end{bmatrix}.
$$
Using Proposition \ref{p5.1.3}, it is easy to calculate that the (output-based) simulations of these two models are the same as
$$
\bar{x}(t+1)=\begin{bmatrix}
0&1\\
1&1\\
\end{bmatrix}
\bar{x}(t).
$$
Hence, system $\Sigma$ is output robust.
\end{exa}

Consider a control network (\ref{5.1.1}) where $u(t)\in \D_{\ell}$, $M\in {\cal L}_{n\times n\ell}$ satisfying (\ref{5.2.2}). Assume there exists a state feedback control
\begin{align}\label{5.2.6}
u(t)=Gx(t),
\end{align}
where $G\in {\cal L}_{\ell\times n}$, such that the closed-loop system is output robust, then $u(t)$ is called an output robust control, which solves the
output robust control problem.

\begin{exa}\label{e5.2.3}
Consider a Boolean control network $\Sigma$ with nominated model as
\begin{align}\label{5.2.7}
\begin{array}{l}
\begin{cases}
x_1(t+1)=(\neg x_1(t))\vee (\neg u(t)),\\
x_2(t+1)=x_1(t)\bar{\vee} x_3(t),\\
x_3(t+1)=\left[x_1(t)\bar{\vee} x_2(t)\right]\vee x_3(t),
\end{cases}\\
~~y(t)=\left[x_1(t)\lra x_2(t)\right]\lra \neg x_3(t);
\end{array}
\end{align}
and its disturbed model as
\begin{align}\label{5.2.8}
\begin{array}{l}
\begin{cases}
x_1(t+1)=(\neg \xi(t))\wedge x_1(t) \wedge u(t),\\
x_2(t+1)=\left[\xi(t)\vee \neg x_1(t)\right]\bar{\vee} x_3(t),\\
x_3(t+1)=\left[x_1(t)\bar{\vee} x_2(t)\right]\vee x_3(t),
\end{cases}\\
~~y(t)=\left[x_1(t)\lra x_2(t)\right]\lra \neg x_3(t).
\end{array}
\end{align}

It is easy to see that if we choose
\begin{align}\label{5.2.9}
u(t)=x_1(t),
\end{align}
then the closed-loop system becomes (\ref{5.2.3})-(\ref{5.2.4}), which is output robust. Thus, the state-feedback control (\ref{5.2.9}) solves the output robust problem of $\Sigma$.
\end{exa}

Output robust control solves the disturbance decoupling problem without the regularity assumption \cite{che11b}.

\section{Conclusion}

This paper investigated the transition representation of BCNs. The main contribution consists of three parts: (i) The topology of TSs was considered, and the formula for calculating fixed points and cycles of BNs was extended to TSs. (ii) Two types of state-based TS representations of BCNs, namely representations with either distinct or non-distinct controls, were proposed. (iii) An output-based TS representation, also called simulation, was studied. Its dynamic equation was obtained. The output robust (control) was also studied.  The technique proposed in this paper is applicable to any finite value network. In fact, some examples in this paper are not BN or BCN. But the technique used for them is exactly the same as the one for BN or BCN.

Some related problems, such as finding the output robust controls, etc., are left for further study.


\end{document}